\documentclass[a4paper,fleqn,usenatbib]{mnras}
\usepackage[T1]{fontenc}
\usepackage{ae,aecompl}
\usepackage{graphicx}	% Including figure files
\usepackage{amsmath}	% Advanced maths commands
\usepackage{amssymb}	% Extra maths symbols
\usepackage{wasysym}
\usepackage{lscape}
\usepackage{natbib}
\usepackage[varg]{txfonts}
\usepackage{url}
\usepackage{xspace}
\usepackage{xcolor}
\usepackage{longtable}
\usepackage{gensymb}
\usepackage{pdflscape}	% Landscape pages
\bibpunct{(}{)}{;}{a}{}{,}
\pdfoutput=1

\newcounter{Rco}
\newcommand{\Ionst}[1]{\setcounter{Rco}{#1}\Roman{Rco}}
\newcommand{\Ion}[2]{\mbox{#1\,{\scriptsize\Ionst{#2}}}}
\newcommand{\Ionw}[3]{\mbox{#1\,{\scriptsize\Ionst{#2}}~$\lambda\,#3$\,\AA}}

\newcommand{\Ionww}[3]{\mbox{#1\,{\scriptsize\Ionst{#2}}~$\lambda\lambda\,#3$\,\AA}}

\newcommand{\loggw}[1]{\mbox{$\log g\hspace{-0.5mm} =\hspace{-0.5mm}  #1$}}

\newcommand{\Teff}{\mbox{$T_\mathrm{eff}$}\xspace}
\newcommand{\Teffw}[1]{\mbox{$\Teff\hspace{-0.5mm} =\hspace{-0.5mm} #1 \,\mathrm{kK}$}}
\newcommand{\ebv}{$E_\mathrm{B-V}$\xspace}

\newcommand{\Msol}{$M_\odot$}
\newcommand{\Rsol}{$R_\odot$}
\newcommand{\Mdot}{$\dot{M}$}

\newcommand{\hwwds}{hot wind white dwarfs\xspace}
\title[Unravelling the ultrahot wind phenomenon]{Unravelling the baffling mystery of the ultrahot wind phenomenon in white dwarfs
}

\author[Reindl et al.]{
Nicole Reindl$^{1}$\thanks{E-mail: nr152@le.ac.uk},
M. Bainbridge$^{1}$,
N. Przybilla$^{2}$,
S. Geier$^{3}$,
M. Prv\'ak$^{4}$,
J. Krti\v{c}ka$^{4}$,
\newauthor
R.~H.~\O stensen$^{5}$,
J. Telting$^{6}$,
K. Werner$^{7}$
\\
$^{1}$Department of Physics and Astronomy, University of Leicester, University Road, Leicester LE1 7RH, UK\\
$^{2}$Institut f\"ur Astro- und Teilchenphysik, Universit\"at Innsbruck, Technikerstr. 25/8, 6020 Innsbruck, Austria\\
$^{3}$Institute for Physics and Astronomy, University of Potsdam, Karl-Liebknecht-Str. 24/25, 14476 Potsdam, Germany\\
$^{4}$Department of Theoretical Physics and Astrophysics, Masaryk University, Kotl\'a\v{r}sk\'a 2, 611 37 Brno, Czech Republic\\
$^{5}$Department of Physics, Astronomy and Materials Science, Missouri State University, Springfield, MO 65897, USA\\
$^{6}$Nordic Optical Telescope, Rambla Jos\'{e} Ana Fern\'{a}ndez P\'{e}rez 7, E-38711 Bre{\~{n}}a Baja, Spain\\
$^{7}$Institute for Astronomy and Astrophysics, Kepler Center for Astro and Particle Physics, Eberhard Karls University, Sand 1, 72076 T\"ubingen, Germany
}
\date{Accepted 2018 October 1. Received 2018 September 23; in original form 2018 September 23}

\pubyear{2018}

\begin{document}
\label{firstpage}
\pagerange{\pageref{firstpage}--\pageref{lastpage}}
\maketitle

\begin{abstract}
The presence of ultra-high excitation (UHE) absorption lines (e.g., \Ion{O}{8}) in the optical spectra of several of the 
hottest white dwarfs poses a decades-long mystery and is something that has never been observed in any other astrophysical 
object. The occurrence of such features requires a dense environment with temperatures near $10^6$\,K, by far exceeding 
the stellar effective temperature. Here we report the discovery of a new hot wind white dwarf, \mbox{GALEX\,J014636.8$+$323615}. 
Astonishingly, we found for the first time rapid changes of the equivalent widths of the UHE features, which are correlated to 
the rotational period of the star ($P=0.242035$\,d). 
We explain this with the presence of a wind-fed circumstellar magnetosphere in which magnetically confined wind shocks heat up 
the material to the high temperatures required for the creation of the UHE lines. The photometric and spectroscopic variability of 
\mbox{GALEX\,J014636.8$+$323615} can then be understood as consequence of the obliquity of the magnetic axis with respect to the rotation axis 
of the white dwarf. This is the first time a wind-fed circumstellar magnetosphere around an apparently isolated white dwarf has been 
discovered and finally offers a plausible explanation of the ultra hot wind phenomenon. 
\end{abstract}


%%%%%%%%%%%%%%%%%%%%%%%%%%%%%%%%%%%%%%%%%%%%%%%%%%
%%%%%%%%%%%%%%%%% BODY OF PAPER %%%%%%%%%%%%%%%%%%

\section{Introduction}
\label{sect:intro}
White dwarfs represent the end product of evolution for the vast majority of all stars. Amongst the hottest white dwarfs, i.e. 
white dwarfs with effective temperatures (\Teff) higher than 60\,kK, there exists a poorly understood -- but 
nevertheless large -- group of objects, which are commonly referred to as hot wind white dwarfs.
Different to ordinary hot white dwarfs, the Balmer\,/\,\Ion{He}{2} lines of the hot wind white dwarfs are 
unusually deep and broad. The majority of the \hwwds exhibits broad absorption features in 
their optical spectra, identified as ultra-high excitation (UHE) absorption lines (i.e., \Ion{C}{5}, 
\Ion{C}{6}, \Ion{N}{6}, \Ion{N}{7}, \Ion{O}{7}, \Ion{O}{8}, \Ion{Ne}{9}, \Ion{Ne}{10}). This is something that has 
never been observed in any other astrophysical object.\\
The occurrence of these obscure features requires a dense environment with temperatures of the order $10^6$\,K, 
by far exceeding the stellar effective temperature. A photospheric origin can therefore be ruled out. Since  
some of the UHE lines often exhibit an asymmetric profile shape, it was suggested that those lines might form 
in a hot, optically thick stellar wind \citep{Werneretal1995}, hence the alternative designation of these stars as 
hot wind white dwarfs. The physical process responsible for this suspected extremely hot, optically thick stellar 
wind, however, poses a decades-long mystery.\\
The fraction of the hottest white dwarfs affected by the ultra-hot wind phenomenon is significant.
Every other H-deficient white dwarf with \Teff$>60$\,kK (corresponding to the DO spectral type) belongs to the 
group of hot wind white dwarfs, and also H-rich white dwarfs are affected by this phenomenon. In total, eighteen 
H-deficient and three H-rich hot wind white dwarfs are currently known (\citealt{Werneretal1995, Dreizleretal1995, 
Werneretal2004, Huegelmeyeretal2006, Werneretal2014, Reindletal2014c}, Kepler et al. in preparation). In the latter the 
UHE features are typically weaker and, thus, more difficult to detect. The majority of the hottest H-rich white 
dwarfs does, however, display the Balmer-line problem \citep{Tremblayetal2011}, which serves as a first indicator 
for the hot wind phenomenon. Thus, it may be speculated that a significant fraction of all stars experiences this 
phenomenon at the beginning of the white dwarf cooling sequence. Therefore, this phenomenon urgently awaits an 
explanation.
\vspace*{-3mm}
\section{Photometric observations}
\label{sect:photo}
\begin{figure}
  \includegraphics[width=\columnwidth]{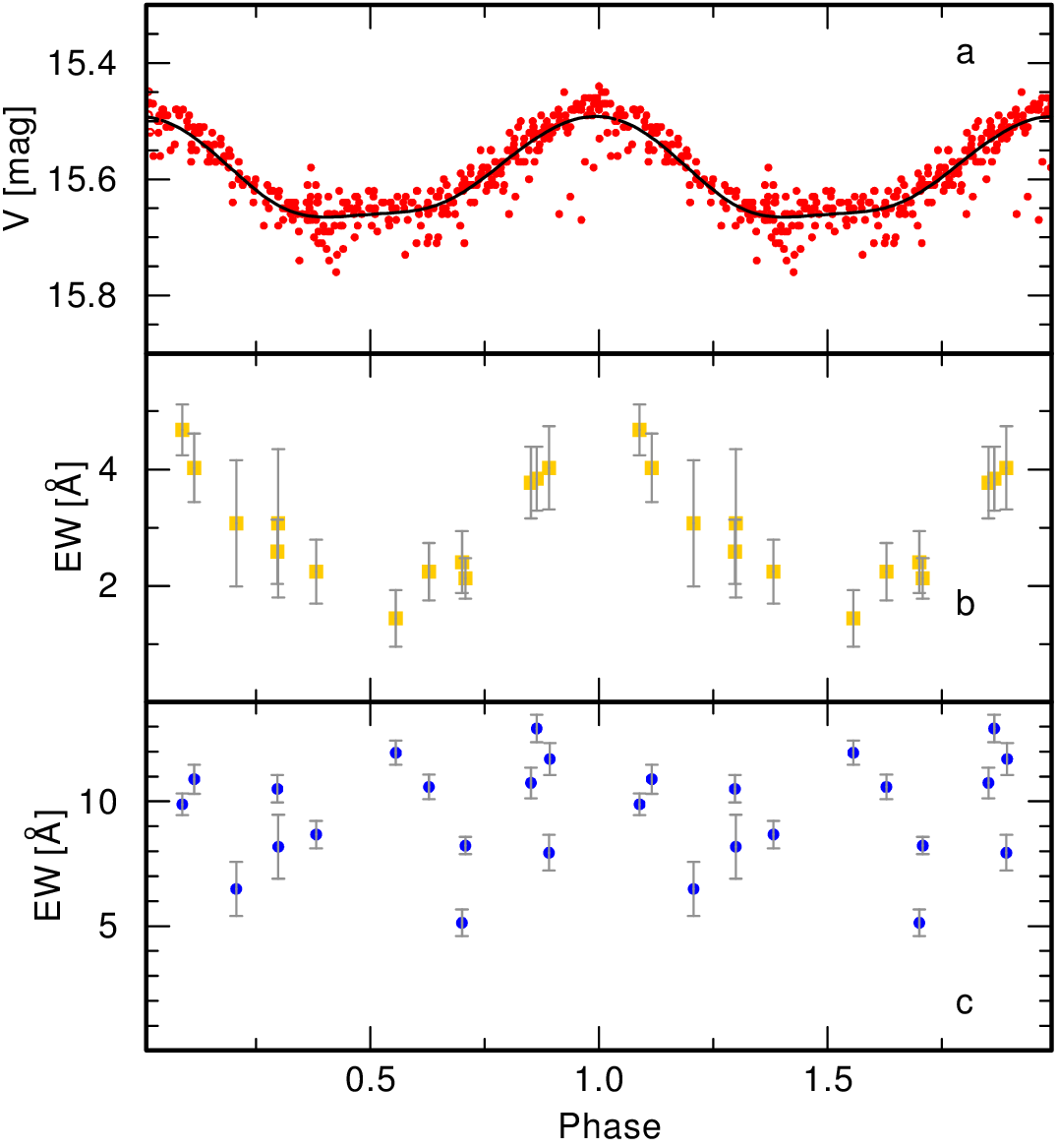}
\vspace*{-5mm}
  \caption{Phase folded light curve of J0146+3236 over-plotted with a bright spot model (a). Panels b and c show the equivalent widths of the UHE feature located at 6060\,$\AA$ (yellow) and \Ionw{He}{2}{6560} (blue), respectively.}
  \label{fig:phase}
\end{figure}

Based on the light curve from the Catalina Sky Survey (CSS) DR1, \cite{Drakeetal2014} classified \mbox{GALEX\,J014636.8$+$323615} 
(from now on J0146+3236) as a post-common envelope binary with a period of $0.484074$\,d and a V-band amplitude of 0.17\,mag. 
We obtained the V-band light curve of J0146+3236 released in the CSS DR2. Since the data points are spaced unevenly, 
we calculated a Lomb-Scargle periodogram (Fig.~\ref{fig:lomb}) as outlined in \cite{Press1989}. We find a period of 
$P=0.242035$\,d with a False Alarm Probability (i.e. the uncertainty associated with the probability that the peak was chosen 
entirely wrong) of $2 \times 10^{-51}$. In Fig.~\ref{fig:phase} the phase folded light curve is shown. Our 
derived period is half of that given by \cite{Drakeetal2014}, however, they required their folded light curve to have two 
minima which may have led to an alias in their period estimate.  
We determine the ephemeris of predicted maxima of the light curve to be \mbox{$\mathrm{HJD}= 2456595.98922 + 0.242035(1)\,E$.}
\vspace*{-3mm}

\begin{figure*}
\includegraphics[width=16.0cm]{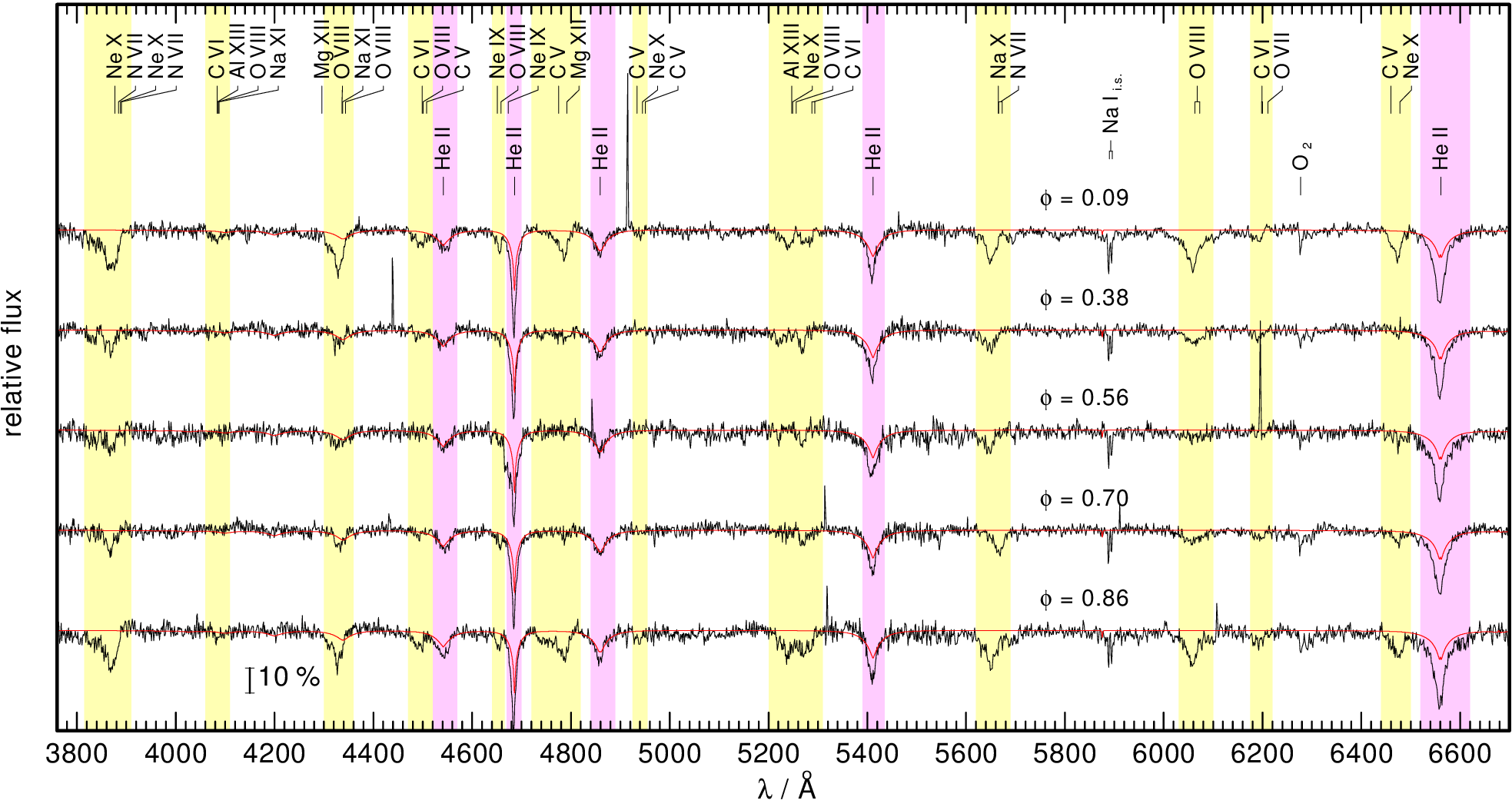}
\vspace*{-3mm}
\caption{TWIN spectra (gray) of J0146+3236. Over-plotted is a pure He TMAP model with \Teffw{100} and \loggw{7.5} (red). He lines are highlighted in pink and UHE features in yellow. Identified lines are marked. The vertical bar indicates 10\% of the continuum flux.}
\label{fig:GJ0146a}
\end{figure*}

\section{Spectroscopic observations}
\label{sect:spec}
We first observed J0146+3236 during a survey exploring candidate sdB stars for the Kepler mission, which were selected 
from the GALEX survey. The spectrum was taken with the Nordic Optical Telescope using the Alhambra Faint Object Spectrograph 
and Camera with grism 16 ($R\approx 1000$). Based on this observation, we originally classified J0146+3236 as a normal DO 
white dwarf. This star was also discovered in the course of the Large Sky Area Multi-Object Fiber Spectroscopic Telescope (LAMOST, 
\citealt{Cuietal2012}) DR3 by \cite{GentileFusilloetal2015}, who also classified J0146+3236 as a DO white dwarf. In October/November 
2014 we performed spectroscopic follow-up at the Calar Alto $3.5$\,m telescope (ProgID H14-3.5-022) using the TWIN spectrograph 
and a slit width of 1.2 acrsec. We used grating No. T08 for the blue channel and No. T04 for the red channel. The spectra have a 
resolution of $1.8\,\AA$ and a typical signal to noise ratio of 50. A total of 14 spectra of J0146+3236 were taken in three consecutive 
nights. After each spectrum, we required ThAr wavelength calibration. The data reduction was done using IRAF. We did not 
flux-calibrate our data.\\
A selection of these spectra along with line identifications is shown in Fig.~\ref{fig:GJ0146a}. For comparison, we over-plotted 
a pure He model spectrum with \Teffw{100} and \loggw{7.5} that was calculated with the T{\"u}bingen non-LTE Model-Atmosphere 
Package (TMAP, \citealt{werneretal2003, tmap2012, rauchdeetjen2003}). The \Ionww{He}{2}{4686, 4859, 5411, 6560} lines in the 
spectra of J0146+3236 appear unusually broad compared to the photospheric DO white dwarf model. The lack of \Ionw{He}{1}{5876} 
(present in the spectra of some hot wind white dwarfs) indicates an effective temperature in excess of $100$\,kK. We note, that 
J0146+3236 is the hitherto brightest known hot wind white dwarf and displays particularly strong UHE absorption lines. 
These lines have been tentatively identified as Rydberg lines of ultra-high excited metals. A unique assignment of the lines to 
particular elements is not possible, although they probably stem from C, N, O, and Ne \citep{Werneretal2018}.
The TWIN spectra of J0146+3236 also allow for the first time the identification of an additional UHE feature located at 4082\,\AA. 
This could be due to \Ionw{C}{6}{4084}, \Ionw{Al}{13}{4084}, \Ionw{O}{8}{4087}, or \Ionw{Na}{11}{4087} 
\citep{Mohretal2008}.\\
Most astonishingly, however, we discover for the first time rapid changes of the line strengths of several UHE features within 
about an hour. The TWIN observations were taken one year after the last data point from the CSS allowing us to derive 
the phase of each spectroscopic observation. We created Voigt profile models of each UHE absorption line using GVPFIT 
\citep{BainbridgeWebb2017} a program based on an "Artificial Intelligence" algorithm, and measured the equivalent widths. 
Interestingly, we find a clear correlation between the change of the equivalent widths of the UHE lines and the light curve 
variability. As can be seen from panel b in Fig.~\ref{fig:phase} the equivalent widths of the UHE lines increase as the object 
becomes brighter, i.e., reaching the maximum at $\Phi\approx 0$, while at $\Phi\approx 0.5$ some of the lines completely vanish. 
We also found variations in the equivalent widths of the \Ion{He}{2} lines, however, these appear randomly as illustrated in panel 
c of Fig.~\ref{fig:phase} for \Ionw{He}{2}{6560}.
\vspace*{-5mm}
\section{Discussion} 
\label{sect:discussion}
\begin{figure}
  \includegraphics[width=\columnwidth]{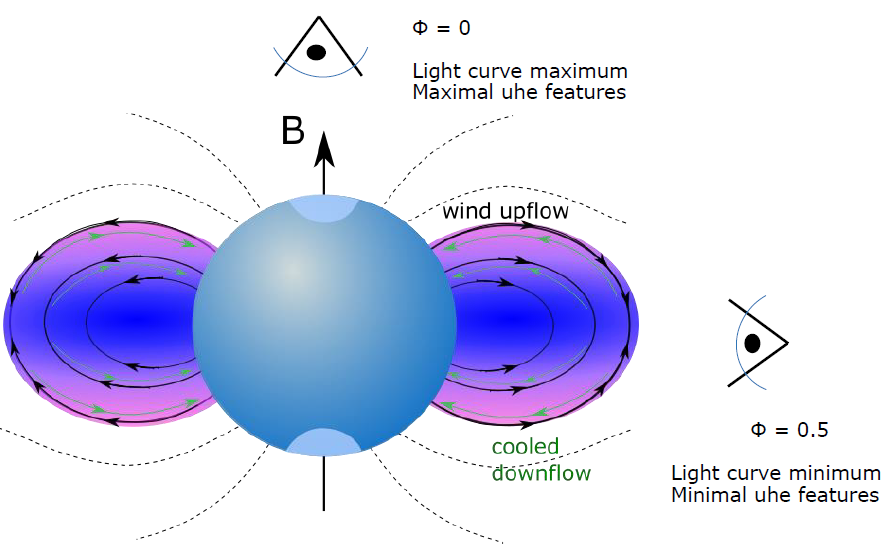}
\vspace*{-5mm}
  \caption{Schematic of J0146+3236 (blue) and its circumstellar environment as seen from the perspective of the star's magnetic axis. Wind material is trapped within closed magnetic loops (black lines) and shock-heated. The UHE lines are produced at high latitudes in the magenta region. This shock-heated plasma then cools and falls back onto the star along closed magnetic field lines (green lines). The rotation axis of the white dwarf is inclined by $45\degree$ to the magnetic axis and also by $\approx 45\degree$ to the observer. At phase $0$, we see the system magnetic pole on, the light curve has its maximum because the entire shock-heated torus and possibly a bright spot are visible. The strengths of the UHE lines is maximal. At phase $0.5$ some of the torus is occulted by the star and the strengths of the UHE features becomes minimal.}
  \label{fig:model}
\end{figure}

The observed photometric period and the amplitude of the light curve are too high to be attributed to pulsations.  
Periods and amplitudes of hot, pulsating white dwarfs (i.e., the GW Vir stars) are of the order of order of minutes and a 
few mmag, respectively \citep{Fontaine2008b}. 
It is also unlikely that the photometric variability is caused by the presence of a close companion for  
several reasons. Firstly, we cannot detect any radial velocity variations of the \Ion{He}{2} absorption lines larger than 
$14.2$\,km/s. Therefore, a possible companion should have either a high inclination angle or a low mass. 
The shape of light curve of J0146+3236 resembles that of \mbox{SDSS\,J212531.92$-$010745.8} 
\citep{Nageletal2006, Schuh2009, Shimansky2015}. This system is composed of hot a H-deficient white dwarf and a close,  
low mass main sequence companion ($P=0.29$\,d). In that case the shape of the light curve is caused by the reflection effect 
of the cool companion. However, in contrast to \mbox{SDSS\,J212531.92$-$010745.8}, in the spectra of J0146+3236 the 
hydrogen Balmer series does not appear in emission as expected due to the irradiation of the cool companion by the hot white dwarf. 
Also an infra-red excess, which could reveal the presence of a possible companion, cannot be detected\footnote{For this, we 
obtained FUV, NUV \citep{Bianchi2014}, u, g, r, i, z \citep{Ahnetal2012}, J, H \citep{Cutri2003}, and W1 \citep{Cutri2014} 
magnitudes of J0146+3236 and converted them into fluxes as outlined in \cite{Reindletal2016, Verbeeketal2014}. We 
applied a reddening of \ebv $=0.05$ \citep{Schlafly2011} to the flux of our TMAP model mentioned above. Thereafter, we 
normalized the reddened model flux to the $g$ magnitude and found that broadband color measurements agree within 
3$\sigma$ with the model flux.}.\\
Interpreting the photometric period as the rotation period of the star, a spot on the surface of the white dwarf 
could explain the photometric variability. Convection can be ruled out as at such high effective temperatures the 
atmospheres are radiative. Hence, this spot must be generated by a magnetic field. Spots have been detected on several 
hot (\Teff$>$30\,kK), magnetic white dwarfs, causing similar amplitudes in the light curves at the rotation period 
of the star \citep{Hermesetal2017b}. Most magnetic white dwarfs are slow rotators (as it occurs for non-magnetic 
white dwarfs, \citealt{Fontaine2008b}), i.e. they show rotation periods of several hours to days 
\citep{Kawaler2004, Hermesetal2017}. We stress, at such a low rotation rate ($v_{\mathrm{rot}}= 1$ km/s, assuming 
a typical white dwarf radius of 0.02\,\Rsol), rotational broadening is not detectable in spectra. The sharp cores of the 
\Ion{He}{2} lines in the spectra of J0146+3236 (and that of all other hot wind white dwarfs) show no evidence of 
Zeeman splitting, suggesting an upper limit on the magnetic field strength of 100\,kG. Note that spots have been 
detected on white dwarfs with even lower magnetic field strengths \citep{Hermesetal2017}.\\
The black line in panel b of Fig.~\ref{fig:phase} shows that the shape of the light curve of J0146+3236 could be explained 
with a bright spot model assuming an inclination of $i = 50\degree$. The synthetic light curve was calculated with a surface 
brightness distribution that provides the best match to the observed light curve as outlined in \cite{Prvaketal2015}.  
Our model assumes a brightness of slightly over 125\% relative to the rest of the stellar surface and the final surface brightness 
distribution is shown in Fig.~\ref{fig:map}. Spots on hot white dwarfs are actually expected to be caused by the accumulation of metals 
around the magnetic poles \citep{Hermesetal2017c}, but such models would predict much lower amplitudes for the light curve 
($<0.1$\,mag for hot white dwarfs, Krti\v{c}ka in preparation). Yet, a magnetic field offers another attractive explanation for the 
photometric variation and the occurrence of the variable UHE features, namely the presence of a wind-fed circumstellar 
magnetosphere.\\
Such magnetospheres develop through the trapping of wind by closed magnetic loops and are commonly observed in magnetic 
O- and B-type main sequence stars \citep{Owockietal2016}. The accelerated wind material, from opposite footpoints of closed magnetic 
loops, collides near the loop apex leading to magnetically confined wind shocks (MCWS, \citealt{Babel1997, udDoula2016}). 
These MCWSs heat up the magnetically-confined plasma (typically located within a few stellar radii of the stellar surface) 
to several MK \citep{udDoula2016, Wade2017}, as required for the creation of the UHE lines observed in the 
hot wind white dwarfs.\\
Magnetohydrodynamic (MHD) simulation studies (e.g., \citealt{udDoula2002, udDoulaetal2008}) show that the overall net effect 
of a large-scale, dipole magnetic field can be characterized by the wind magnetic confinement parameter
$\eta_{\mathrm{\star}}=\frac{B_{\mathrm{eq}}^2 R_{\mathrm{\star}}^2}{\dot{M}_{B=0} v_\infty},$
where $B_{\mathrm{eq}}$ is the field strength at the magnetic equatorial surface radius $R_{\mathrm{\star}}$, 
and \Mdot$_{B=0}$ and $v_\infty$ are the fiducial mass-loss rate and terminal wind speed that the star would have in the absence 
of any magnetic field. For $\eta_{\mathrm{\star}} > 1$, outflow near the magnetic equator is trapped within the Alfv\'{e}n radius 
by closed magnetic loops, forming a wind-fed circumstellar magnetosphere. If we assume for J0146+3236 a typical white dwarf radius of 
$R_{\mathrm{\star}} =0.02$\,\Rsol, a mass of $M=0.6$\,\Msol, and \Teff$=100$\,kK, then the mass loss rate is expected to be 
$\dot{M}_{B=0} = 10^{-11}$\Msol/yr and the terminal wind velocity about $10^4$\,km/s \citep{UnglaubBues2000}. In that case 
the requirement for a wind-fed circumstellar magnetosphere is already fulfilled for $B_{\mathrm{eq}}>0.6$\,kG (well below the 
upper limit of 100\,kG drawn from the absence of Zeeman splitting).\\
The periodic variation of the UHE lines and the light curve can then be understood with the following scenario which we illustrate 
in Fig.~\ref{fig:model}. The star is represented in blue, the spots at the magnetic poles in light-blue, and the circumstellar magnetosphere 
in blue, purple, and magenta. The density of the magnetosphere is expected to increase towards lower latitudes, while the 
temperature is expected to decrease starting with $\approx10^6$ to $10^7$\,K at high latitudes (origin of the UHE lines) and reaching 
the photospheric \Teff\ near the magnetic equator \citep{Owockietal2016}. Let us assume that the magnetic dipole field is tilted, with respect 
to the rotation axis, by $\approx 45\degree$. At phase $0$ we see the system magnetic pole-on and the maximum amount of the 
hot post-shock gas located at high latitudes. Therefore, the strengths of the UHE lines is maximal. Also the light curve has its maximum, 
as the entire shock-heated torus and possibly a bright spot are visible. As the star rotates the torus becomes partly occulted by the star, 
the projected area of the torus decreases, and the cooler parts of the magnetosphere come to the fore. Thus, the light curve diminishes, 
reaching its minimum phase $0.5$ when we see the system edge-on. Also the strengths of the UHE features decreases 
as less of hot post-shock gas from the high latitudes of magnetosphere is visible in the edge-on view.\\
We note that the variability of the UHE lines in J0146+3236 is similar to what is observed in \mbox{$\theta$ Ori C}, a O-type main 
sequence star hosting a circumstellar magnetosphere and whose magnetic axis is also tilted by $\approx 45\degree$ with respect to the 
rotation axis (see e.g., \citealt{Gagne2005}). Phase resolved Chandra grating observations of \mbox{$\theta$ Ori C} show that the line 
fluxes of H- and He-like ions are approximately 30\% higher when the system is seen pole-on than when seen edge-on \citep{Gagne2005}.
The observation of UHE absorption lines in the spectra of J0146+3236 suggests that the shock-heated plasma must be optically thick, 
in contrast to the optically thin shock-heated plasma observed in \mbox{$\theta$ Ori C}, which shows the X-ray UHE lines in emission. 
X-ray UHE absorption lines have been detected in warm absorbers which are observed in about 50\% of the Seyfert galaxies 
(e.g., \citealt{Blustinetal2002}). The column densities of these warm absorbers which surround the active nucleus of these galaxies 
are of the order of $10^{22}$\,cm$^{-2}$ \citep{Komossa2000}. The magnetospheric column density of the shock-heated plasma of 
\mbox{$\theta$ Ori C} is slightly lower, namely $5 \times 10^{21}$\,cm$^{-2}$ \citep{Gagne2005}. The characteristic wind density, 
$\rho_w \equiv \dot{M}_{B=0} / 4\pi v_\infty R_{\star}^2 \approx 3\times10^{-13}\mathrm{g/cm^3}$, of J0146+3236 is similar to  
of \mbox{$\theta$ Ori C}. 
We speculate the higher magnetospheric density of J0146+3236 might be related to the fact that low luminosity stars with weak winds 
experience weaker shocks \citep{ud-Doulaetal2014}. The slower scaled shock speed, $\omega_s$, leads to a cooler post-shock temperature 
$T_s \sim \omega_s^2$ \citep{Owockietal2016} and the shock-heated plasma becomes more spatially extended. The subsonic nature of the 
post-shock flow implies a nearly constant pressure ($P\sim\rho T$), meaning that the density increases strongly as temperature declines 
(see Fig.~3 in \citealt{Owockietal2016}). Hence, we would expect for the hot wind white dwarfs a higher density and lower temperature 
magnetosphere (of the order $10^6$\,K instead of $10^7$\,K) than what is observed for O- and B-type main sequence stars that host 
circumstellar magnetospheres.\\
The cooler parts of the magnetosphere then likely constitute an additional line forming region of the too-broad and too-deep 
\Ion{He}{2} lines. \cite{Werneretal2018} already speculated these lines could be caused by relatively cool gas in a 
circumstellar static cloud. Similarly, variable optical helium lines, like \Ionw{He}{2}{4684}, in magnetic O-stars show clear 
signatures of being formed in a dynamical magnetosphere \citep{Grunhutetal2012, Wadeetal2015}. \cite{udDoula2002} showed that 
the shock-heated wind plasma eventually cools radiatively back to temperatures near the stellar effective temperature and then 
falls back on to the stellar surface along closed magnetic field lines on a dynamical (free-fall) timescale. The infall of the 
cooled material occurs in sporadic intervals of highly compressed, localized streams \citep{Owockietal2016}, which may explain 
the random variations of the equivalent widths of the \Ion{He}{2} lines.\\
Having in mind that the UHE features diminish when we see the system edge-on, this scenario also naturally explains 
why some objects only show abnormal broad and deep optical H and \Ion{He}{2} lines, but no UHE lines. 
Another possibility is that these stars have an even slower scaled shock speed and, thus, a cooler post-shock plasma. 
A similar argument may hold for H-rich white dwarfs, amongst which the ultra-hot wind phenomenon seems less common. H-rich 
objects have lower $v_\infty$ compared to their H-deficient counterparts \citep{Pauldrachetal1988}, therefore a lower 
post-shock temperature ($T_\mathrm{s} \sim v_\infty^2$) would be expected. We also note that the variability of the UHE lines 
is only expected if the magnetic and rotation axis are tilted.
\vspace*{-7mm}
\section{Summary and conclusions} 
\label{sect:conclusions}
We report about the discovery of a new hot wind white dwarf, \mbox{J0146+3236}, that reveals rapid changes 
of the equivalent widths of several UHE features, which are correlated to the star's rotation period ($P=0.242035$\,d). 
We present a physical model that, finally, offers a plausible explanation of the ultra-hot wind phenomenon. A weak magnetic 
field ($\approx 0.6 - 100$\,kG) channels the wind material toward the magnetic equator, forming a dense, wind-fed circumstellar 
magnetosphere. MCWSs heat up the material to the high temperatures required for the creation of the UHE lines. As the star's 
magnetic axis (i.e., the axis of the torus) is tilted with respect to the rotation axis, the brightness and intensity of 
the UHE features vary as a function of the rotation phase of the star. The radiatively cooled and dense post-shock material in the 
magnetosphere sporadically falls back onto the stellar surface and constitutes an additional line forming region for the variable,
too-broad and too-deep \Ion{He}{2} lines.\\
This is the first time that a wind-fed circumstellar magnetosphere around an apparently isolated white dwarf has been discovered. 
About 25\% of all white dwarfs have weak magnetic field strengths of a few kG \citep{Cuadrado2004}, 
suggesting that indeed a substantial fraction of all stars may develop a wind-fed circumstellar magnetosphere during their 
early white dwarf cooling phase. We like to emphasize that this discovery may have far-reaching consequences. The neglect of 
the magnetospheric contribution to the optical H and \Ion{He}{2} lines could lead to large systematic effects on the 
derived effective temperatures and surface gravities of these stars, and consequently on all other properties obtained from those
(e.g., the mass distributions and luminosity functions of hot white dwarfs). In addition wind magnetic spin-down 
\citep{Ud-Doulaetal2009} could have a noticeable impact on the angular momentum evolution of these stars.\\
Further theoretical and observational effort is therefore highly desirable for a profound understanding of the hot wind white dwarfs. 
Spectro-polarimetric observations may help to derive the magnetic field strengths of the hot wind white dwarfs. 
MHD simulations and a detailed line profile analysis of the UHE features would allow us to investigate the physics of the shock heated plasma. 
Time-resolved, high S/N UV spectra (that is where photospheric metals can be detected) could reveal chemical spots and possible 
signatures of a weak, phase dependent wind in the UV resonance lines. Detailed light curve modelling could help to understand 
the origin of the photometric variability more and reveal the geometry of these extraordinary systems.
\vspace*{-7mm}
\section*{Acknowledgements}
Based on observations collected at the German-Spanish Astronomical Center, Calar Alto, jointly operated by the 
Max-Planck-Institut f\"ur Astronomie Heidelberg and the Instituto de Astrof\'{i}sica de Andaluc\'{i}a (CSIC).
We thank JJ Hermes for helpful discussions and comments.
NR is supported by a Royal Commission 1851 research fellowship. MB is supported by a Leverhulme Trust Research Project Grant.
SG acknowledges funding by the Heisenberg program of the Deutsche Forschungsgemeinschaft under grant GE 2506/8-1.
MP and JK were supported by grant GA\,\v{C}R  16-01116S. 
This research used the ALICE High Performance Computing Facility at the University of Leicester. 
The CSS survey is funded by the National Aeronautics and Space
Administration under Grant No. NNG05GF22G issued through the Science Mission Directorate Near-Earth Objects 
Observations Program. The CRTS survey is supported by the U.S.~National Science Foundation under grants AST-0909182.
\vspace*{-7mm}
\bibliographystyle{mnras}
\bibliography{hotwind}
%%%%%%%%%%%%%%%%%%%%%%%%%%%%%%%%%%%%%%%%%%%%%%%%%%
%%%%%%%%%%%%%%%%%%%%%%%%%%%%%%%%%%%%%%%%%%%%%%%%%%
%%%%%%%%%%%%%%%%%%% APPENDICES %%%%%%%%%%%%%%%%%%%%%

\appendix
\section{Online material}

\begin{figure}
  \includegraphics[width=\columnwidth]{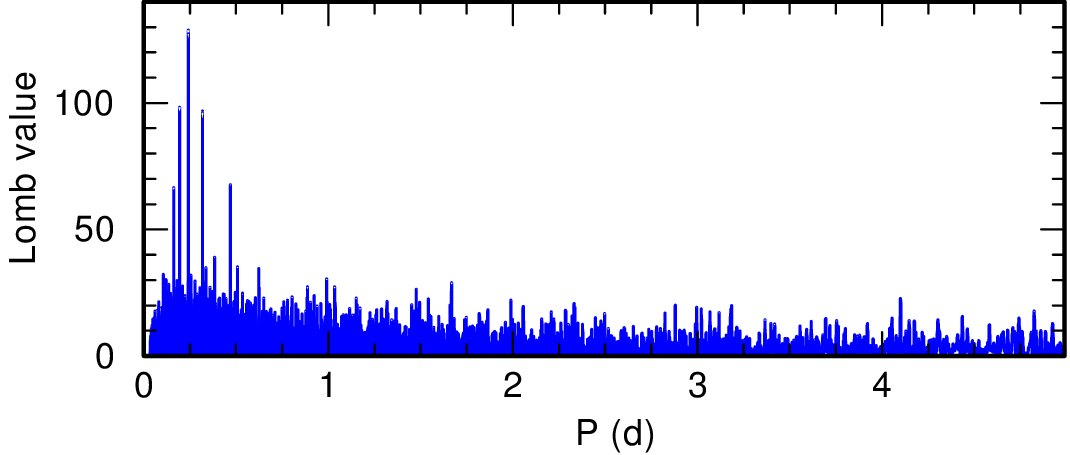}
  \caption{Lomb-Scargle periodogram of the CSS DR2 light curve of J0146+3236.}
  \label{fig:lomb}
\end{figure}

\begin{figure}
  \includegraphics[width=\columnwidth]{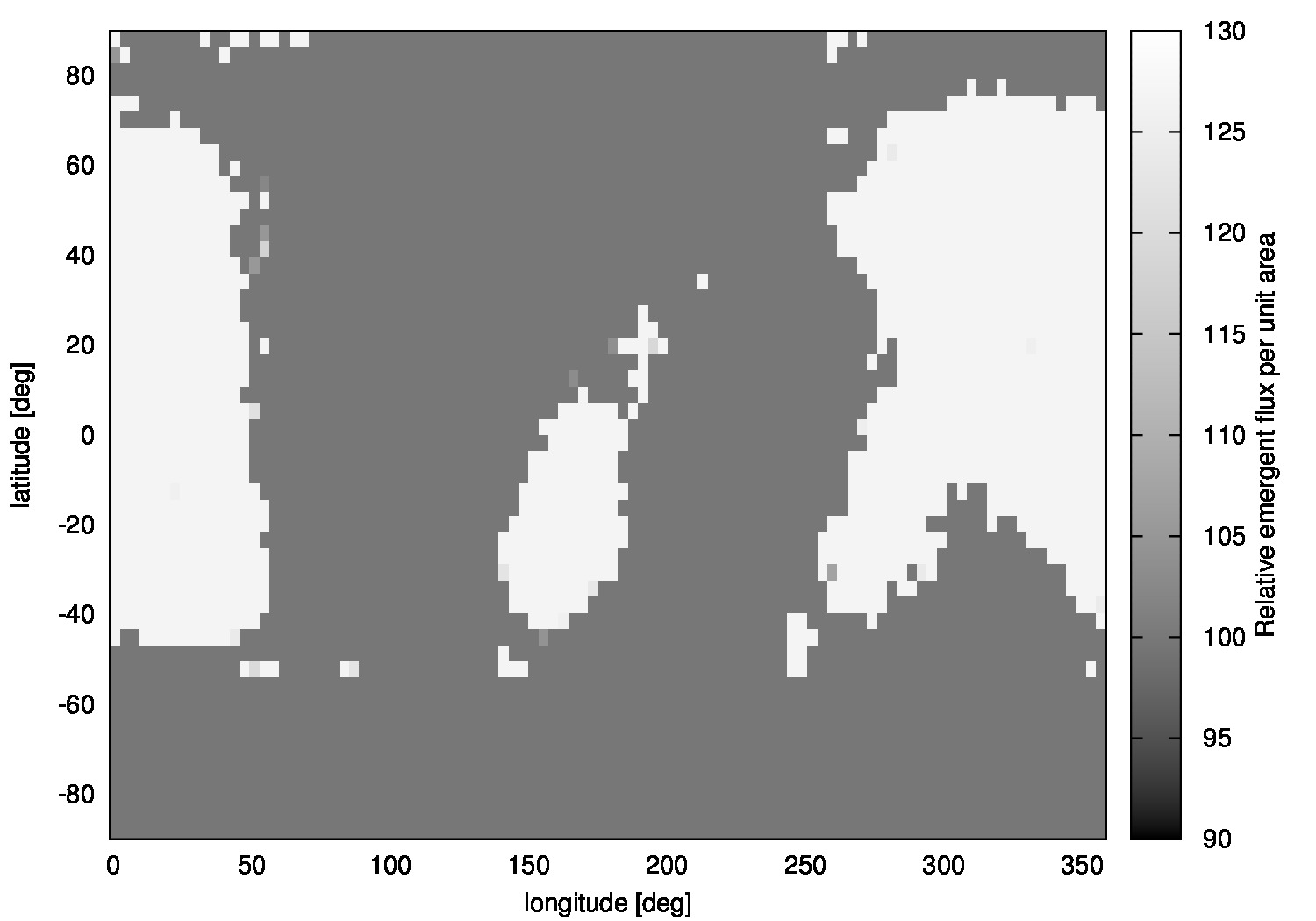}
%\vspace*{-5mm}
  \caption{Emergent flux map of our spot model for J0146+3236.}
  \label{fig:map}
\end{figure}
% Don't change these lines
\bsp	% typesetting comment
\label{lastpage}
\end{document}